\documentclass{article}
\usepackage[letterpaper, total={6in, 8in}]{geometry}

\usepackage{xr-hyper}
\usepackage{amsmath}

\usepackage[colorlinks,urlcolor=blue,linkcolor=blue,citecolor=blue]{hyperref}
\usepackage{cite}
\usepackage{multicol}
\usepackage{multirow}

\makeatletter
\newcommand*{\addFileDependency}[1]{
  \typeout{(#1)}
  \@addtofilelist{#1}
  \IfFileExists{#1}{}{\typeout{No file #1.}}
}
\makeatother

\newcommand*{\myexternaldocument}[1]{%
    \externaldocument{#1}%
    \addFileDependency{#1.tex}%
    \addFileDependency{#1.aux}%
}
\myexternaldocument{sup}

\usepackage{comment}
\usepackage[export]{adjustbox}
\usepackage{ifthen}
\usepackage{amssymb}
\usepackage{xcolor}
\usepackage{comment}

\newboolean{showcomments}
\setboolean{showcomments}{true}

\makeatletter
\newcommand{\mynote}[3]{%
  \ifthenelse{\boolean{showcomments}}{%
   \fbox{\bfseries\sffamily\scriptsize#1}%
   {\small$\blacktriangleright$\textsf{\emph{\color{#3}{#2}}}$\blacktriangleleft$}}%
  {%
   \@bsphack
   \@esphack
  }%
}
\makeatother

\definecolor{asparagus}{rgb}{0.53, 0.66, 0.42}

\newcommand{\so}[1]{\mynote{SO}{#1}{red}}

\usepackage{multirow}
\usepackage{multicol}
\usepackage{amsmath}
\usepackage{algorithm}
\usepackage[noend]{algpseudocode}
\usepackage[caption=false,font=footnotesize,labelfont=sf,textfont=sf]{subfig}
\usepackage{threeparttable}
\usepackage{siunitx}

\DeclareSIUnit\carboneq{CO{_2}eq}

\begin{document}


\title{Sustainable AI Processing at the Edge}

\author{Sébastien Ollivier, Sheng Li, Yue Tang, Chayanika Chaudhuri, Peipei Zhou,\\ Xulong Tang, Jingtong Hu, and Alex K. Jones\\
University of Pittsburgh}



\markboth{Special Issue on Environmentally Sustainable Computing}{Sustainable AI Processing at the Edge}
\maketitle

\begin{abstract}
Edge computing is a popular target for accelerating machine learning algorithms supporting mobile devices without requiring the communication latencies to handle them in the cloud.  Edge deployments of machine learning primarily consider traditional concerns such as SWaP constraints (Size, Weight, and Power) for their installations.  However, such metrics are not entirely sufficient to consider environmental impacts from computing given the significant contributions from \textit{embodied} energy and carbon.  In this paper we explore the tradeoffs of convolutional neural network acceleration engines for both inference and on-line training.  In particular, we explore the use of processing-in-memory (PIM) approaches, mobile GPU accelerators, and recently released FPGAs, and compare them with novel Racetrack memory PIM.  Replacing PIM-enabled DDR3 with Racetrack memory PIM can recover its embodied energy as quickly as 1 year.  For high activity ratios, mobile GPUs can be more sustainable but have higher embodied energy to overcome compared to PIM-enabled Racetrack memory.

\end{abstract}

\section{Introduction} 
\label{sec:introduction}

Deep neural networks have become a popular algorithm for a variety of applications using mobile devices including smart phones but also recently expanding to connected and autonomous vehicles (CAVs), robotics, or even unmanned aerial vehicles (UAVs), and other smart infrastructure.  Convolutional Neural Networks (CNNs) have been demonstrated to provide solutions to these problems with relatively high accuracy.  While there have been many proposals to improve the performance and energy efficiency of CNN inference, these algorithms are too compute and data intensive to execute directly on mobile nodes typically operating with limited computational and energy capabilities.  Thus, edge servers, now being deployed often in conjunction with advanced (\textit{e.g.,} 5G) wireless networks, have become a popular target to accelerate CNN inference.  Moreover, due to their deployment in the field, edge servers must operate under size, weight, and power (SWaP) constraints, while serving many concurrent requests from mobile clients.  Thus, to accelerate CNNs, these edge servers often use energy-efficient accelerators, reduced precision, or both to achieve fast response time while balancing requests from multiple clients and maintaining a low operational energy cost.  Recently, there has been a trend to push online training to edge server nodes to avoid communicating large datasets from edge to cloud servers~\cite{yue2022}.  However, online training typically requires much higher precision and floating-point computation compared to inference.

Unfortunately, the proliferation of computing, both the mobile devices, and the edge servers themselves, can come at the expense of negative environmental impacts.  Broadly, there are many concerns from computing infrastructure that range from the use of scarce rare earth elements to environmental costs from extracting materials for manufacturing integrated circuits and energy storage (i.e., batteries).  Furthermore, computing infrastructure can lead to problematic emissions of everything from carcinogens to volatile organic compounds, not to mention green-house warming gases (GWG), particularly CO$_2$, but also including methane and others.  As such, there is a significant and growing aspect of environmental impacts that come from embodied impacts of computing.  Embodied impacts include the energy, carbon emissions, etc. from manufacturing computing infrastructure and particularly the semiconductor elements that form the heart of all computing systems.  Recent evidence shows that for cloud servers, embodied impacts are equally as high as operational (run-time) effects~\cite{dell-server-lca2019}.  For mobile devices and compact computers, embodied impacts can reach 80-90\% of total life-cycle impacts and that these impacts are dominated by their integrated circuits~\cite{Jones-ICCAD-Sus-2013,dell-server-lca2019,Dark-Silicon-Harmful}.  Thus, for SWaP optimized systems, embodied energy is a higher proportion of the energy footprint making its amortization an important goal.    

While specialty processing units including field-programmable gate arrays (FPGAs) and graphics-processing units (GPUs) can accelerate CNN applications while meeting low operational energy constraints, they come at the cost of increasing the silicon area of these edge systems.  This creates a significant tradeoff between embodied energy from including accelerators and the operational energy impacts from executing the algorithms.

In this paper we explore our recently proposed processing-using-memory proposal of Racetrack memory (RM) to implement both CNN inference and training and compare the results with state-of-the-art proposals using GPU, FPGA, and PIM using DRAM.  Our comparison considers the main two phases of energy consumption of embodied and operational energy~\cite{Jones-ICCAD-Sus-2013}.  Thus, we explore its total energy efficiency compare to other state-of-the-art technique to implement modern applications.  This allows a evaluation of the sustainability of the system choices.  We select energy as our metric as it bridges the manufacturing and operational phase of the system into a metric that can be directly compared.  However, we will also discuss how these energy values inform other environmental metrics including GWG when including electrical grid mix profiles.

In particular, this paper makes the following contributions:
\begin{itemize}
    \item We provide estimates of the embodied energy to fabricate PIM-enabled domain wall memory and recent GPU and FPGA comparison points.
    \item We characterize the operational power and performance of representative CNN applications for edge-scale execution including both inference and training.
    \item We conduct indifference and brake-even analyses of different target systems and usage scenarios to determine holistic sustainability calculations.
    \item We explore the environmental impacts of these systems.
\end{itemize}

In the next section we discuss the background and related work to conduct these analyses.

\section{Background}

The primary source of environmental impacts for computing systems comes from the integrated circuits that implement the core functionality of processing, data storage, etc~\cite{Jones-ICCAD-Sus-2013}.  To determine the holistic environmental impacts in terms of energy, GWG, and other concerns of a product or process, such as semiconductor fabrication, typically involves a technique called Life Cycle Assessment (LCA)~\cite{bilec1}.  LCA is most accurate when a detailed analysis of the process is used to determine the assessment, but sometimes relative costs to similar processes can be used as a coarse-grain assessment called economic input/ouput (EIO) LCA.  Relatively few process LCAs have been undertaken of semiconductor fabrication.  One assessment considered CMOS, Flash, and DRAM fabrication covering technologies from \SI{350}{\nano\meter} down to \SI{32}{\nano\meter}~\cite{boyd2011life}.  Hybrid LCA used used process technology trends and EIO LCA to create a model scaling to \SI{7}{\nano\meter}~\cite{Dark-Silicon-Harmful}.  Recently, a similar effort with additional process information was used to estimate technologies from \SI{28}{\nano\meter} to  \SI{3}{\nano\meter}~\cite{bardon2020dtco}.  This study includes a detailed analysis of the impact of moving from deep-ultraviolet (DUV) lithography to extreme ultraviolet (EUV), which relieves aggressive multiple patterning requirements for sub \SI{30}{\nano\meter} features.

\subsection{Indifference and Break-even analyses}
\label{sec:indifference}

One motivation to use a single metric of energy for both manufacturing and operational sustainability evaluation of the system is to allow quantitative comparison metrics such as indifference and break-even analyses. To compare two design choices to select the appropriate system for deployment we can use the indifference formula $t_I$ as shown in Eq.~\ref{eq:indifference}~\cite{KLINE2019322}.  For a system with higher embodied energy ($M$) and lower operational energy ($P$), $t_I$ is the time at which the increase in embodied energy will be completely amortized by the savings in operational energy.  Thus, if the proposed service time $t<t_I$ the architecture with the lower embodied energy minimizes environmental impact.  In contrast, for a proposed service time $t>t_I$ the architecture with the lower operational energy minimizes impact.  If one choice is lower in both embodied and operational energy, then indifference analysis is not needed and the lower energy system can be selected independent of service time.  A similar calculation can be considered for the break even time $t_B$, also defined in Eq.~\ref{eq:indifference}~\cite{KLINE2019322}.  Consider the case that an existing system is already deployed.  Replacing the existing system is like assuming embodied energy of the deployed system is 0.  Thus, $t_B$ is the time it takes for the replacement system to overcome the embodied energy of the replacement through operational energy savings.  $t_B = t_I$ when $M_0 = 0$.

\begin{equation}
t_I = \frac{M_1-M_0}{P_0-P_1} \qquad \qquad t_B=\frac{M_1}{P_0-P_1}
\label{eq:indifference}
\end{equation}



While we characterize several accelerators in this work for CNN acceleration, we also consider an exotic technology that uses spintronics to store data and has been explored for PIM called Racetrack memory~\cite{9045991}.  We provide some background on RM in the next section.

\subsection{Racetrack Memory}

\begin{figure}[tbp]
\centering
\includegraphics[width=.5\columnwidth]{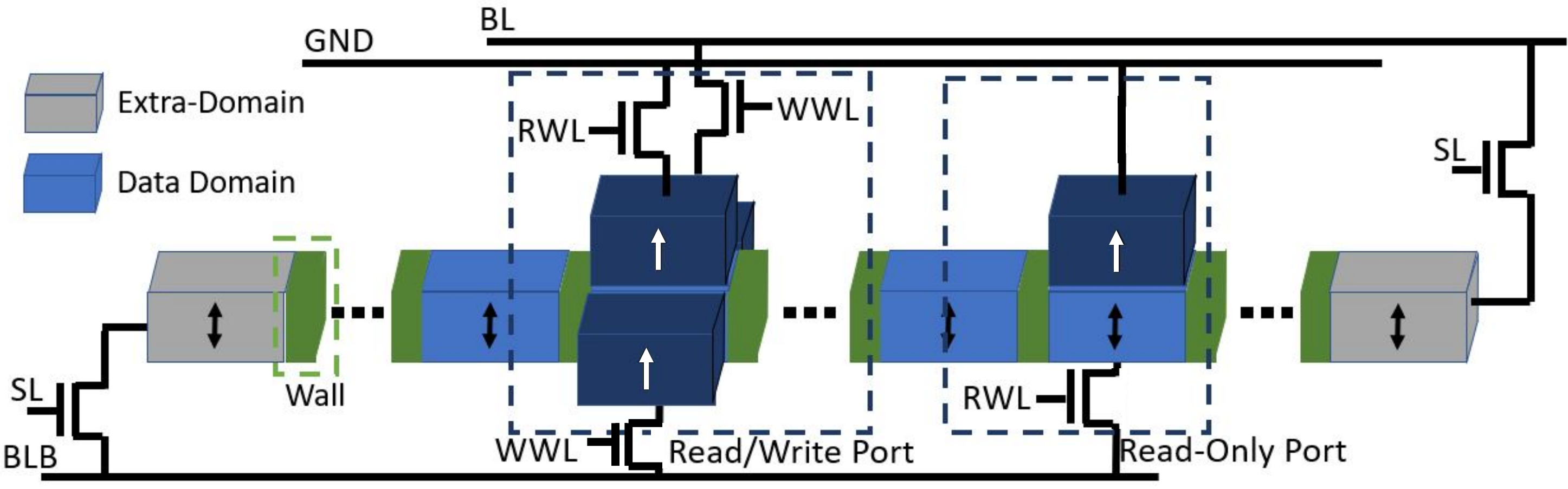}
\caption{Anatomy of a domain-wall memory nanowire. }
\label{DWMzoom}
\label{fig:nanowire}
\vspace{-.2in}
\end{figure}

Spintronic RM is made of ferromagnetic nanowires consisting of many magnetic domains separated by domain walls (DWs) as shown in Fig.~\ref{DWMzoom}. Each domain has its own magnetization direction  
such that binary values are represented by the magnetization direction of each domain, either parallel/antiparallel to a fixed reference. For a planar nanowire, several domains share one/few access point(s) (APs) for read and write operations~\cite{zhang2012perpendicular}. RM is similar to and has many of the same advantages as STT-MRAM, including high endurance, fast access time, low energy, particularly static energy due to the device's non-volatility.   RM can have a density $\leq$2F$^2$ because it can store multiple bits in a nanowire accessed using one transistor.  In contrast, STT-MRAM requires 6-50F$^2$.  Hence, RM has been proposed at several memory levels, from L1 cache to main memory.  RM achieves this density by requiring shifting if data is not aligned with the access point.  Shifting occurs through DW motion in the nanowire.  DW motion is controlled by applying a short current pulse laterally along the nanowire governed by \texttt{SL}. Random access requires {\em shifting} the target domain to {\em align} it with an AP (dark blue) and apply a current to {\em read} or {\em write} the target bit.  To avoid data loss when shifting, the blue domains store actual data while the grey domains are overhead domains to prevent data loss. 
Shift-based writing (Read/Write Port)~\cite{DWM_Tapestri} allows slower current writes to be replaced with orthogonal shifts from fixed magnetic alignment domains to reduce latency and energy.

RM, like many other novel memories including resistive memory PIM crossbars in PRIME, has also received significant attention for PIM, particularly for deep learning~\cite{yu2014energy,PIRM}.

\subsection{Convolutional Neural Networks}

\newcommand{\myMatrix}[1]{\mathbf{{#1}}}
CNNs are a popular method to compute deep learning algorithms.  CNNs are dominated by the convolution operation, which is a windowed point-wise multiplication accumulation of multiple channels of input features with a set of weights to generate output features.  As an example, for the input features $\myMatrix{I}$ and weights $\myMatrix{K}$ of size $N\times R_{in}\times C_{in}$ and $M\times N\times 3\times 3$, respectively, the convolution operation for the window at $m$ (output channel index), $r$ (row), $c$ (column) is:

\footnotesize \[
Conv(\myMatrix{I},\myMatrix{K})(m,r,c)= \sum_{n=0}^{N-1}\sum_{j=0}^{2} \sum_{t=0}^{2}  \myMatrix{K}_{m,n,j,t}\times\myMatrix{I}_{n,r+j,c+t}
\vspace{-.05in}
\] \normalsize\\
where $M$ is the number of output channels, $N$ is the number of input channels, $R_{in}\times C_{in}$ is the size of an input feature map.

While deep learning with CNNs presumes calculations with floating-point values, CNN inference calculations can often be reduced to integer computation with as few as 8-bits achieving reasonable accuracy.  Recent DRAM PIM work has shown that in many cases this can be further reduced to ternary $w\in\{-1,0,1\}$ or even binary $w\in\{0,1\}$ computations operations to replace the multiplications.  However, online training for all but the simplest CNNs still requires full 32-bit floating-point computations to work properly.  Without this accuracy, the weight updates can be ineffective and possibly even detrimental.

In the next section we explore embodied energy calculations of a variety of accelerators suitable for CNN acceleration.

\section{Evaluation of Edge Acceleration Sustainability}
\label{Sec:Results}

To consider holistic energy across embodied and operational phases requires LCA of the semiconductor fabrication process.  In the next section we discuss how to obtain embodied energy and carbon footprint.

\subsection{Determining Embodied Energy and Carbon}
\label{sec:fabrication}

RM, like STT-MRAM, is a spintronic memory that can be implemented by adding additional layers of ferromagentic materials and insulators on top of the completed CMOS layers.  Typically these are added in between the lower levels of the metal stack.  The spintronic devices are composed of three conceptual layers, a fixed magnetic layer, an MgO barrier that separates the fixed layer from the free layer often made out of a ferromagentic material like a CoFeB nanowire.  CoFeB with different doping properties can also be used for fixed magnetic layers.  In terms of the process steps, they are essentially the same between STT-MRAM and RM.  Thus, during manufacture, in addition to the CMOS and metal layers, three additional mask layers are required due to the different substances of each DWM sub layer. They are composed of three lithography, three dry etching and a deposition step~\cite{STT-MRAM-Embodied}.  We used a modified version of NVSIM~\cite{dong2012nvsim} to calculate the die area of the NVM.  We then calculated the additional die area required to support the PIM units area from PIRM~\cite{PIRM}. 

To determine the embodied energy of the DRAM, FPGA, and GPU we used the die area and technology node along with process LCA reported in the literature for \SI{350}{\nano\meter}--\SI{32}{\nano\meter}~\cite{boyd2011life} processes and for \SI{28}{\nano\meter}--\SI{3}{\nano\meter}~\cite{bardon2020dtco}.  There is a significant gap between the two studies with a significant gap between the reported \SI{32}{\nano\meter}~\cite{boyd2011life} and \SI{28}{\nano\meter}~\cite{bardon2020dtco} such that a third study that reports \SI{32}{\nano\meter}~\cite{5156786} sits between the two.  Thus, in our work we do not compare nodes that cross the studies.


\begin{table}[tbp]
    \centering
    \caption{Energy to \SI{}{\gram \carboneq \per {\kilo\watt\hour}}~\cite{mai-eere-grid} and Grid Mixes~\cite{NYTIMES}}
    \label{tab:grid-mixes}
    \begin{tabular}{l|c||c|c|c|c}
    \hline\hline
    \multicolumn{2}{r||}{\textbf{Source} \textbf{\SI{}{\gram \carboneq \per{\kilo\watt\hour}}}} & \textbf{AZ} & \textbf{CA} & \textbf{TX} & \textbf{NY}\\\hline
         Coal & 980 & 20\% & 3\% & 19\% & --\\
         Natural Gas & 465 & 40\% & 39\% &53\%& 37\%\\\hline
         Geothermal & 27 & -- & 5\% & -- & --\\
         Hydroelectric & 24 & 5\% & 18\% & -- & 22\%\\
         Solar PV & 65 & 7\% & 20\% &2\%& 2\% \\
         Wind & 11 & -- & 7\% &17\%& 4\% \\
         Nuclear & 27 & 28\% & 7\% &9\%& 33\%\\
         Biopower & 54 & -- & 3\% & -- & --\\\hline
         \multicolumn{2}{l||}{\textbf{Mix} (\SI{}{\gram\carboneq \per{\kilo\watt\hour}})} & 395 & 234 & 438 & 188\\\hline\hline
    \end{tabular}
\end{table}
Prior to reporting the embodied energy we created studied several grid mix scenarios reported in Table~\ref{tab:grid-mixes}.  Using the carbon footprint of multiple different electrical generation methods we report the grid mix for Arizona, California, Texas, and New York, all of which have significant semiconductor fabrication activity and very different grid mixes.    Arizona and Texas have significant electrical generation from coal and the highest generation from natural gas.  While Arizona has significant generation from nuclear plants and Texas has significant wind energy, their 395 and 438 \SI{}{\gram} of CO$_2$ equivalent generated per \SI{}{\kilo\watt\hour} are much higher than California and New York, which still get more than a third of their electricity from natural gas.  California is very balanced on renewable energy and New York has significant hydroelectric and nuclear power generation, thus their grid mix generates about half the carbon at 234 and 188\SI{}{\gram \carboneq  \per{\kilo\watt\hour}}, respectively.  

\setlength{\tabcolsep}{4pt}
\begin{table}[tbp]
\centering
\caption{Accelerator statistics, embodied energy, and embodied carbon emissions for grid mixes from Table~\ref{tab:grid-mixes}.}
\label{tab:Sustainability}
\begin{threeparttable}
\begin{tabular}{|l|c|c||c||c|c|c|}
\hline
 & RM & DDR3 & RM & RM & FPGA & GPU  \\
\hline
\hline
Tech Node & 32\footnotemark[1]\textsuperscript{,}\footnotemark[4] & 55\footnotemark[1] & 32\footnotemark[2]\textsuperscript{,}\footnotemark[4] & 32\footnotemark[3]\textsuperscript{,}\footnotemark[4] & 7\footnotemark[3] & 14\footnotemark[3]\\
\hline
Die Size (\SI{}{\milli\meter\squared}) & 38 & 73 & 38 & 38
 & 324 & 350 \\
\hline
Die per wafer & 1847 & 967 & 1847 & 1847 & 217 & 201 \\
\hline
PE (\SI{}{\kilo\watt\hour\per Wafer}) & 1626 & 1200 & 1254 & 832 & 1482 & 882 \\
\hline\hline
Energy (\SI{}{\mega\joule\per die}) & 3.17 & 4.47\footnotemark[5] & 2.44 & 1.62 & 24.59 & 15.80 \\
\hline\hline
AZ (\SI{}{\gram \carboneq \per die}) & 348 & 490\footnotemark[5] & 268 & 178 & 2698 & 1734 \\\hline
CA (\SI{}{\gram \carboneq \per die}) & 206 & 291\footnotemark[5] & 159 & 105 & 1598 & 1027  \\\hline
TX (\SI{}{\gram \carboneq \per die}) & 386 & 544\footnotemark[5] & 297 & 197 & 2992 & 1922\\\hline
NY (\SI{}{\gram \carboneq \per die}) & 166 & 233\footnotemark[5] & 127 & 85 & 1284 & 825 \\\hline
\end{tabular}
\begin{tablenotes}
\footnotesize 
\item[1] Calculated using process LCA from~\cite{boyd2011life}.
\item[2] Calculated using process LCA from~\cite{5156786}.
\item[3] Calculated using process LCA from~\cite{bardon2020dtco}.
\item[4] Requires extra steps for spintronics~\cite{STT-MRAM-Embodied}.
\item[5] Requires 16 dies to build a the tested 1GB DIMM.
\end{tablenotes}
\end{threeparttable}
\end{table}
\setlength{\tabcolsep}{6pt}
In Table~\ref{tab:Sustainability} we report the embodied energy and embodied carbon using the grid mixes from Table~\ref{tab:grid-mixes}.  The RM is augmented with PIM capabilities to compute, binary, ternary, integer, and floating point computation~\cite{PIRM,FPIRM}.  We calculated the sustainability DDR3-1600 as this is the device that has been used to implement DRAM PIM using ELP$^2$IM~\cite{elp2im} and conduct ternary model reduction of CNN inference.  For dedicated accelerators we selected edge server appropriate low-energy devices including the Versal Prime FPGA (VM1802) from AMD/Xilinx and the NVIDIA Jetson NX mobile GPU.  Note, we were somewhat limited in our choice of, particularly FPGA, devices as die area is not typically reported, which is necessary to determine embodied energy/carbon estimates.  We note that the RM is extremely dense, even compared to the DDR.  However, the GPU and FPGA require an order of magnitude more embodied energy due to their much larger die sizes.  

\subsection{Holistic Sustainability Evaluation}

\setlength\tabcolsep{2pt}
\begin{table}[tbp]
\centering
\caption{Performance, Operational Power, and Efficiency per Power and Carbon of Different Edge Accelerators}
\label{tab:CNNResult}
\begin{tabular}{l|c|c|c|c|c}
\hline
\multicolumn{6}{c}{\textbf{Inference Acceleration using Ternary Model Reduction and PIM}}\\\hline
\textbf{Benchmark} & \textbf{Target} &  \textbf{Throughput}  & \textbf{Power} & \multicolumn{2}{c}{\textbf{Efficiency}} \\
 & & FPS & W & FPS/W & MF/gCO$_2$eq\\
\hline
\hline
\textbf{Alexnet} & DDR3~\cite{elp2im} & 84.8 & 2 & 42.4 & 0.35--0.81\\
Ternary~\cite{elp2im} & RM & 490 & 0.93 & 526 & 4.6--10.8\\\hline
\hline
\multicolumn{6}{c}{\textbf{Training Acceleration using Floating-Point 32 Data}}\\\hline
\textbf{Benchmark} & \textbf{Target} &  \textbf{Throughput}  & \textbf{Power} & \multicolumn{2}{c}{\textbf{Efficiency}} \\
 & & GFLOPS & W & GFLOPS/W & TFLOPS/gCO$_2$eq\\
\hline
\multirow{3}{*}{\textbf{Alexnet}} 
& GPU & 1335 & 21.05 & 63.4 & 521--1214 \\
& RM & 50.72 & 5.65 & 8.97 & 74--172 \\
& FPGA &  34.52 & 7.74 & 4.46 & 37--85 \\
\hline
\multirow{3}{*}{\textbf{VGG-16}} 
& GPU & 848 & 20.37 & 41.6 & 342--797 \\
& RM & 81.95 & 5.7 & 14.37 & 118--275\\
& FPGA & 46.99 & 7.71 & 6.09 & 50--117\\

\hline

\end{tabular}
\vspace{-.2in}
\end{table}

\begin{figure*}[tbp]
\centering
\subfloat[AlexNet inference PIM breakeven]{
\includegraphics[height=1.775in]{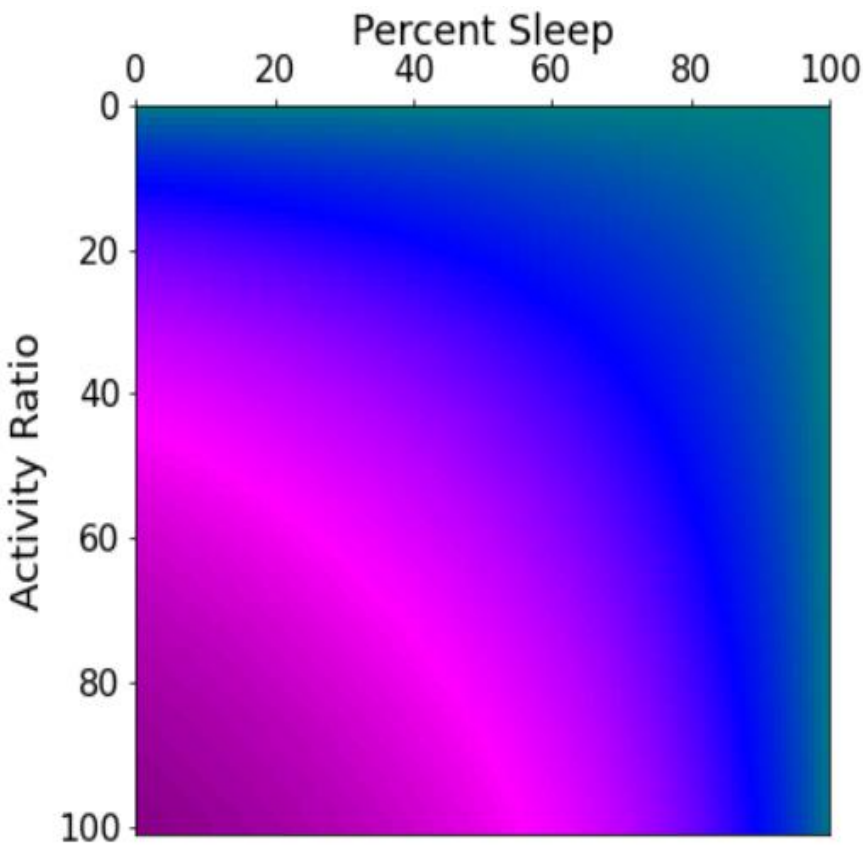}
\label{fig:AlexNet-PIM-Breakeven}
}
\subfloat[AlexNet training GPU/RM indiff.]{
\includegraphics[height=1.775in]{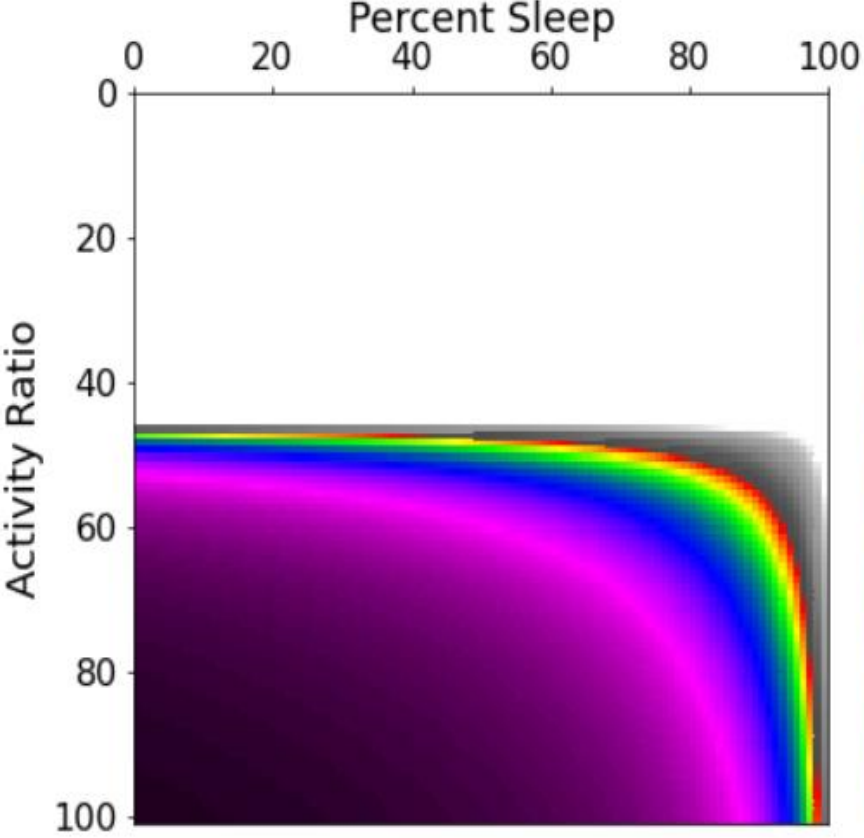}
\label{fig:AlexNet-Training-Indifference}
}
\subfloat[VGG-16 training GPU/RM indiff.]{
\includegraphics[height=1.775in]{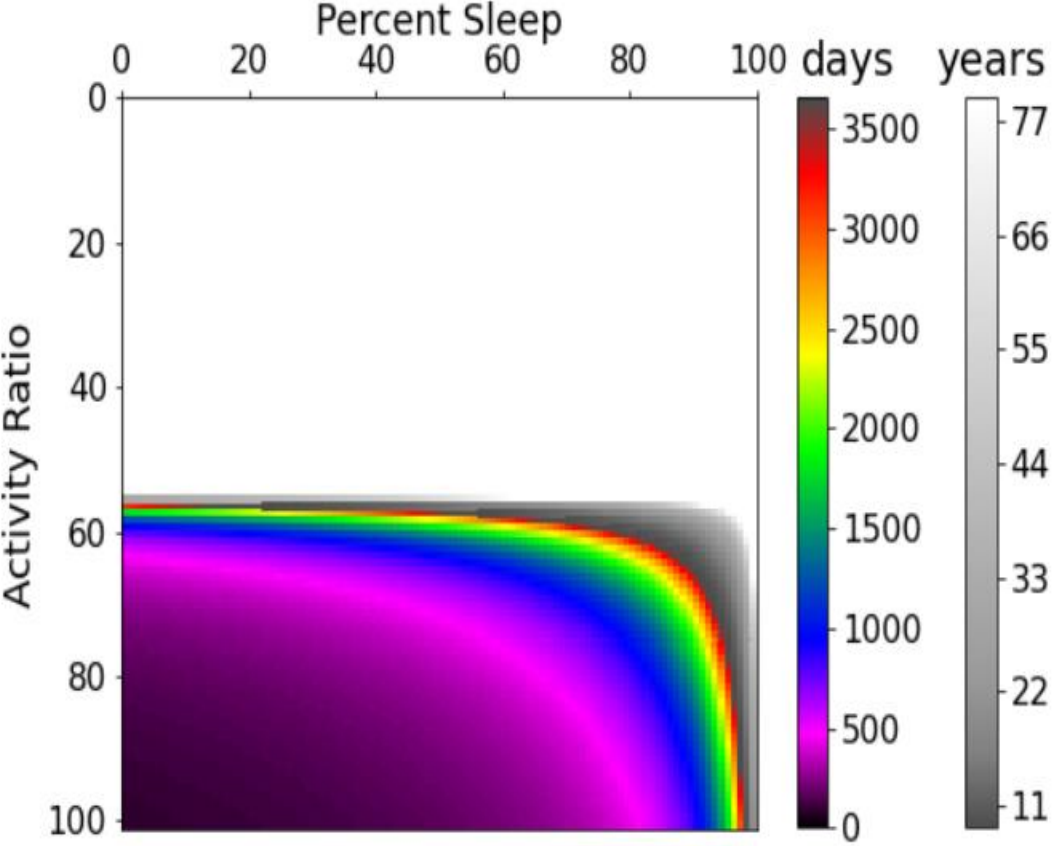}
\label{fig:VGG-16-Training-Indifference}
}
\caption{Sustainability analyses of different accelerator choices for edge systems.}
\label{fig:sus-edge}
\end{figure*}

To determine the overall energy (and carbon footprint) of these acceleration choices we compared a CNN conducting inference using DRAM PIM~\cite{elp2im} and RM PIM~\cite{PIRM}.  RM provides both an embodied and operational energy improvement, ultimately providing order-of-magnitude benefits in mega frames per \SI{}{\gram \carboneq }.  Of course presuming the edge system already contains DDR3, an important experiment is to evaluate the breakeven time beyond which adding the RM acceleration would holistically improve the energy.  We illustrate this using the GreenChip tool~\cite{KLINE2019322} in Figure~\ref{fig:AlexNet-PIM-Breakeven}.  The chart shows the \textit{activity ratio}, which is the ratio of compute to idle time, and the \textit{sleep ratio} which is the ratio of active to sleep time~\cite{KLINE2019322}.  Note, in this case the accelerator could be idle or sleeping while the host is still used.  If the accelerator is active and operating nearly constantly the break-even time can be as low as 1 year.  However, as the usage drops to 50\% the break-even time moves to around 500 days, with lower usage scenarios pushing into the 2--3 time frame and beyond, reaching around 4 years in the upper right corner.  

To explore CNN training we compare the GPU and FPGA using a PyTorch-based flow with hand optimization to generate GPU and FPGA implementations of the AlexNet and VGG-16 training~\cite{yue2022} and hand mapped design for the RM accelerator~\cite{FPIRM}.  From Tables~\ref{tab:Sustainability} and \ref{tab:CNNResult} both embodied and operational energy for the FPGA are higher than both the RM and the GPU, so the indifference calculation will never pick the FPGA.  However, the RM has a lower embodied energy and a higher operational energy than the GPU.  We see the indifference results in Figures~\ref{fig:AlexNet-Training-Indifference} and \ref{fig:VGG-16-Training-Indifference}. For AlexNet in high usage scenarios the indifference time is relatively short to makeup the embodied energy, but starting around 50\% the indifference time starts increasing dramatically such that it becomes impractical in the mid 40\% range.  VGG-16 follows a similar trend, but because the operational energy gap between the two is closer it becomes harder to amortize the embodied energy so the reasonable indifference times fall off sooner.

\section{CONCLUSION}

In this work, we evaluated using RM, which requires additional embodied energy to manufacture, its holistic energy comparison when integrated in edge system while performing CNN applications. The break-even point to replace DRAM PIM with RM PIM resulted in a benefit within the $1\leq t\leq2$ years, likely on the low end of that time-frame given the rising popularity of CNN acceleration on edge servers. In our indifference comparison between RM and the Jetson Xavier NX mobile GPU the edge server activity ratio needs to be at least 40\% for lightweight CNNs like Alexnet and higher for more complex ones like VGG-16 to make a GPU lower overall energy than RM with an in service time of $\leq$10 years.  We also include carbon emissions for both embodied and operational energies where the ranges correspond to the ranges enumerated in the different grid mixes.  Note, the break-even and indifference charts for carbon are identical the same if a consistent grid mix is used.  

Another concern previously raised that remains true is that embodied effects remain high compared to operational effects.  Given it takes tera or even petaflops of GPU compute to be equivalent to a \SI{}{\gram \carboneq } and the amortization time of embodied carbon measured in multiple \SI{}{\kilo\gram \carboneq } is still high (years to decades) and solutions to bring this down are necessary.  We suggest that RM PIM can be a solution insomuch as it competes in the operational phase with order of magnitude lower embodied carbon.

Unfortunately, the observation that the process LCA models are considerably disjoint at their meeting point of a similar technology node is a concern.  Thus, the process LCA still requires improvement and validation, or at least calibration, which makes convergence on a raw magnitude difficult.  However, if this is effectively a constant offset, then the relative comparisons are still extremely useful in allowing considerable design space exploration.

\section{ACKNOWLEDGMENT}

This work was supported in part by the NSF under grants CNS-1822085, CNS-2133267, the National Security Agency, and Laboratory of Physical Sciences.  

\bibliography{pim}
\bibliographystyle{IEEEtran}

\end{document}


\sptitle{Department: Head}
\editor{Editor: Name, xxxx@email}

\title{POD-RACING: Supplementary Material}

\author{Sébastien Ollivier, Xinyi Zhang, Yue Tang, Chayanika Choudhuri, Jingtong Hu,\\ and Alex K. Jones}
\affil{University of Pittsburgh}



\markboth{Special Issue on Artificial Intelligence at the Edge}{POD-RACING}


\maketitle

\section{POD-RACING: Supplementary}

The process to compute integer (or fixed-point) computation using POD-RACING is shown in Algorithm~\ref{alg:int-operations}.  As explained in the main document, addition of $\text{TRD}-2$ operands may be computed directly activating each nanowire in sequence.  For nanowire $N_{i}$, $i$ represents the nanowire that corresponds to a particular bit position of data in a memory row (see Fig.~\ref{subfig:DBC}).  $S_i$, computed using the TR of $N_i$ and the CIM Unit (Fig.~\ref{subfig:PIMUnit}) is written to AP$_0$ of $N_i$ ($S_i \rightarrow N_{i,0}$) where $N_{i,j}$ refers to $N_i$ at AP $j$.  Similarly $C_i \rightarrow N_{i+1,1}$ and $C'_i \rightarrow N_{i+2,0}$.  

The \textsc{Add}$(O[1..5],w)$\footnote{$l\neq0$ can be specified to allow an offset, but we explain for $l=0$.} function of Algorithm~\ref{alg:int-operations} describes this in more detail.  For each nanowire in sequence, starting with the least significant bit, the $S_i$ is computed using the \texttt{XOR} and written to $O[0]_i$, which corresponds to $N_{i,0}$.  Given we have $ones_i$ from the SAs such that $\sum_{k=0}^6 O[k] > i \implies ones_i = 1$ otherwise $ones_i=0$ we can compute the result using multiplexers such that $M_0$ selects $ones_0$ or $ones_2$ based on $ones_1$, $M_1$ selects $ones_4$ or $ones_6$  based on $ones_5$ and $S$ selects $M_0$ or $M_1$ based on $ones_3$.  Similarly, $C$ selects $ones_1$ or $ones_5$ based on $ones_3$  and writes the value in $O[6]_{i+1}$, or $N_{i+1,1}$.  $C'$ is $ones_3$ which is written to $O[0]_{i+2}$ or $N_{i+2,0}$.  By $i=2$ all values of $O[0..6]_{i+2}$ may contain data.  As the most significant bit is reached, values are not written above $N_{w-1}$.  Moreover, just as $S,C,C'$ can be written to $N_i$, $N_{i+1}$, and $N_{i+2}$, simultaneously, when $k$ values are packed into a row, we can activate $k$ nanowires such that $\forall \{0..k-1\} N_{i+k*w}$ generates and writes $S,C,C'$ in parallel.

If it is necessary to add more than $\text{TRD}-2$ numbers, we use \textsc{CSA-Reduction}$(O[0..6])$.  \textsc{CSA-Reduction} is akin to a Carry-Save Adder (CSA) with more than 3 inputs.  A CSA adder computes the $S$ and $C_{OUT}$ from three inputs (\textit{e.g.,} $A,B,C_{IN}$) in parallel before doing a traditional add of $S$ and $C_{OUT}$, which requires traversing the carry chain only once.  Similarly \textsc{CSA-Reduction} computes $\forall i$ $S_i$ in bulk-bitwise fashion, followed by $C_i \rightarrow N_{i+1}$, concluding with $C'_i \rightarrow N_{i+2}$.  While \textsc{Add} takes $w$ cycles to sum five operands, \textsc{CSA-Reduction} takes a constant time to reduce seven operands to three operands.  Only once there are five or fewer remaining operands is \textsc{Add} called.  

We see this use in \textsc{Multiply}$(O_A,O_B,w)$.  First, each bit of $O_A$ is used to determine whether a shifted copy of $O_B$ is copied as a partial product of the multiplication.  We leverage the connections from $N_i \rightarrow \{N_{i-1},N_{i+1}\}$ in Fig.~\ref{subfig:PIMUnit} to accomplish this along with the predication registers~\cite{9065566} shown in Fig.~\ref{subfig:subarray}.  Operand $OP_A$ is logically shifted left while $OP_B$ is logically shifted left so that the predication register which causes a shifted copy of $OP_B$ to be retained can be pulled from the zeroth bit of the rowbuffer to simplify the logic that selects the predication bit source.  In this instance the least significant bit of each packed operand must have a connection to the predication register.   After preparing the partial products they are reduced using \textsc{CSA-Reduction} until five or fewer values remain.  These values are summed using an \textsc{Add}.

\algdef{SE}[VARIABLES]{Variables}{EndVariables}
    {\algorithmicvariables}
    {\algorithmicend\ \algorithmicvariables}
\algnewcommand{\algorithmicvariables}{\textbf{Global variables}}

\definecolor{forestgreen}{RGB}{34, 139, 34}
\algnewcommand{\LeftComment}[1]{\(\triangleright\)#1} 
\algrenewcommand{\algorithmiccomment}[1]{{\color{forestgreen}\(\triangleright\)#1}}

\algrenewcommand\algorithmicindent{1.0em}%

\begin{algorithm}[tbp]
\footnotesize
\caption{Integer/Fixed-point Multiply and Add~\cite{PIRM}}
\label{alg:int-operations}
\begin{algorithmic}[1]

\State \Comment{Fixed-point multiply}
\Function{Multiply}{$O_A,O_B,w$}

\For{$i \gets 0:w-1$} \Comment{w-bit operands}
    \State $P[i] \gets 0$ \Comment{Initialize Partial Product}
    \State \Comment{Predicated copy on $O_A[i]$ of $O_B << i$}
    \State $O_A[0]$ ? $P[i] \gets $\textsc{Copy} $O_B$ 
    \State $O_A \gets O_A >> 1$ \Comment{align $i+1$ with 0 position}
    \State $O_B \gets O_B << 1$ \Comment{prepare $M_B << i$} 
\EndFor
$t \gets \lceil w/7 \rceil; r \gets w\mod7 $
\While {$(t > 1) || (r>5)$}
\For{$i \gets 0:t-1$} 
    \State $P[3i...3i+2]\gets$ \textsc{CSA-Reduction}($P[7i..7i+6]$)
    \If {$r>3$} 
      \State $P[3t...3t\!+\!2]\!\!\gets$\textsc{CSA-Reduction}($P[7t..7t\!+\!r]$)
     \Else 
       \State $P[3t...3t+r] \gets $ $P[7t..7t+r]$)
     \EndIf
     \EndFor
$t \gets  \lceil (t*3+r) / 7 \rceil; r \gets (t*3+r) \mod 7$
\EndWhile
    \State $P[0] \gets $ \textsc{Add}($P[0..r]$,w*2)
    \State \Return{P[0]}
\EndFunction

\State

\State \Comment{Reduce 7 operands of bit-width $w$ to a Sum ($S$), Carry ($C$), and Super Carry ($C'$)}
\Function{CSA-Reduction}{$O[0...6]$}
    \State \Comment{Intrinsic CIM operations allowed using TR~\cite{roxy2020novel}}
    \State \Comment{All bits $i$ in width $w$ in parallel~\cite{PIRM}}
    \ForAll{$i$ in $w$}
        \State $S_i \gets O[0]_i \oplus O[1]_i \oplus ... \oplus O[6]_i$
        \State $ones_i \gets \sum_{k=0}^6 O[k]_i$
        \State $C_i \gets 2 \leq  ones_i < 4 \; \text{\texttt{OR}} \; ones_i \; \geq 6$
        \State $C'_i \gets ones_i \geq 4$
    \EndFor
    \State \Return{$S,C,C'$}
\EndFunction

\State

\State \Comment{Add 5 operands of bit-width $w$ starting at lower-bound bit $l$ (default $=0$), using parallel carry and super carry chains~\cite{PIRM}, and return the Sum ($S$)}
\Function{Add}{$O[1..5],w,l=0$}
    \State \Comment{All bits $i$ in sequence to form carry chain}
    \State \Comment{$O[0],O[6]$ reserved for $C',C$; $S$ overwrites $O[0]$}
    \State $u = l + w$
    \For{$i \gets l:u-3$}
        \State $O[0]_i \gets O[0]_i \oplus O[1]_i \oplus ... \oplus O[6]_i$
        \State $ones_i \gets \sum_{k=0}^6 O[k]_i$
        \State $O[6]_{i+1} \gets 2 \leq  ones_i < 4 \; \text{\texttt{OR}} \; ones_i \; \geq 6$
        \State $O[0]_{i+2} \gets ones_i \geq 4$
    \EndFor
    \State \Comment{Compute $S,C$ for penultimate bit, $S$ for final bit}
    \State $O[0]_{u-2} \gets O[0]_{u-2} \oplus O[1]_{u-2} \oplus ... \oplus O[6]_{u-2}$
    \State $ones_{u-2} \gets \sum_{k=0}^6 O[k]_{u-2}$
    \State $O[6]_{u-1} \gets 2 \leq  ones_{u-2} < 4 \; \text{\texttt{OR}} \; ones_{u-2} \; \geq 6$
    \State $O[0]_{u-1} \gets O[0]_{u-1} \oplus O[1]_{u-1} \oplus ... \oplus O[6]_{u-1}$
    \State \Return{$O[0]$} \Comment{$O[0]$ is the final sum}
\EndFunction
\end{algorithmic}
\end{algorithm}

\section{Floating Point Two-Operand Multiplication}
\label{Sec:FPMult}

As described in the main document, to multiply two floating point operands we follow the flow in Algorithm~\ref{alg:fp-mult}.  First the mantissas of both operands are extracted and reconstructed to include the implicit `1' using bulk bitwise operations as shown in lines~\ref{line:mantissa-mask}--\ref{line:mantissa-mask-end}.  Calling \textsc{Multiply} from Algorithm~\ref{alg:int-operations} the product mantissa is calculated.  The normalization process checks the 48th bit (line~\ref{line:normalization}) and uses this to populate the predication register for a predicated shift operation required to normalize mantissas $M \geq 2.0$ (line~\ref{line:pred-normalize}).  This requires that the predication register must be able to select between bit 0 for multiplication or 47 from the row-buffer to populate the predication register.  

To continue we mask off the exponents (lines~\ref{line:exp-mask}--\ref{line:exp-mask-end}).  Next the new exponent must be calculated, we sum the two extracted exponents $E_A, E_B$ but must subtract (add negative) 127, which is loaded into the memory location indicated by $offset$, to deal with the exponent normalization factor.  If the mantissa was shifted for normalization, we increase the exponent by one through a predicated storage of 0x800000 into an otherwise empty location denoted as $one$.  Executing the \textsc{Add} function sums the exponents $E_A,E_B$, 0x7F800000 (-127), and optionally 0x800000 as a normalization if the mantissa required normalization.  With a 5th 0x0 operand, the new exponent is calculated by adding $w=8$-bits starting at bit $l=23$ (line~\ref{line:exponent-add}).  Note, if $E_{31}$ is `1' the exponent has gone out of range (overflow) because either $E > 255$ or $E < 0$.  In lines~\ref{line:sign}--\ref{line:sign-end} the sign is similarly extracted and computed using an \texttt{XOR}.  Finally in $M, E,$ and $S$ are returned, but the mantissa remains using 47-bits.

\begin{algorithm}[tbp]
\footnotesize
\caption{Floating-point Multiplication}
\label{alg:fp-mult}
\begin{algorithmic}[1]
\State \algorithmicvariables: $OP_A, OP_B, M, E, S$
\State  \Comment {Inputs: $OP_{A,B}$ Operands}
\State  \Comment {Outputs: $M$ = Mantissa, $E$ = Exponent, $S$ = Sign}
\State
\Function{FPMultiply}{$OP_A,OP_B$}
\State \Comment {Mask off and multiply the mantissas}
\label{line:mantissa-mask}
\State $M_A \gets (OP_A$ \texttt{AND} 0x7FFFFF$)$ \texttt{OR} 0x800000
\State $M_B \gets (OP_B$ \texttt{AND} 0x7FFFFF$)$ \texttt{OR} 0x800000
\label{line:mantissa-mask-end}
\State $M \gets$ \textsc{Multiply} $(M_A,M_B)$
\State \Comment{normalize if $M \geq 2.0$}
\State $norm \gets (M$ \texttt{AND} 0x800000000000$)$ 
\label{line:normalization}
\State \Comment{Predicated $M$ normalization shift testing $norm$}
\State $norm$ ? $M \gets M >> 1$ 
\label{line:pred-normalize}
\State \Comment{Mask off, add the exponents with normalization} 
\label{line:exp-mask}
\State $E_A \gets (OP_A$ \texttt{AND} 0x7F800000$)$
\State $E_B \gets (OP_B$ \texttt{AND} 0x7F800000$)$
\label{line:exp-mask-end}
\State $offset \gets$ 0xC0800000 \Comment{~-127 to correct exponent}
\State $one \gets$ 0x0
\State $norm$ ? $one \gets $ 0x800000 \Comment{increase exponent if $M\!\geq\!2.0$}
\State \Comment{Add exponents, note overflow if $E_{31} = 1$}
\State $E \gets $ \textsc{Add}$(E_A,E_B,offset,one,$0x0,8,23)
\label{line:exponent-add}
\State \Comment{Mask off and determine final sign}
\label{line:sign}
\State $S_A \gets (OP_A$ \texttt{AND} 0x80000000$)$
\State $S_B \gets (OP_B$ \texttt{AND} 0x80000000$)$
\State $S \gets (S_A$ \texttt{XOR} $S_B)$
\label{line:sign-end}
\State \Return {$M,E,S$}
\EndFunction
\end{algorithmic}
\end{algorithm}

\section{Floating-Point Multi-Operand Addition}
\label{Sec:FPAdd}

As presented in the main document we conduct floating-point addition following the flow in Algorithm~\ref{alg:fp-add}.  The function begins by searching for the maximum exponent among the operands in groups of TRD$=7$ seven (lines~\ref{line:find-max-exponent}--\ref{line:find-max-exponent-end}) using \textsc{FindMax} from the helper functions in Algorithm~\ref{alg:fp-add-helper} as follows: Starting with the most significant bit ($p$) an \texttt{OR} operation ($isAnyOne$) detects any `1's at that position (line~\ref{line:test-ones}).  If there is at least one `1' then we eliminate values with `0's at that position because they will be smaller by resetting the rowbuffer and writing them back as 0x0.  The algorithm does this by reading each exponent in turn from AP$_0$, inverts the exp with an \texttt{XOR} with 1 ($nIsOne$) then \texttt{AND}s this with $isAnyOne$ (lines~\ref{line:test-local-one}--\ref{line:test-local-one-end}).  A predication bit $pred$ is extracted from the row buffer at position $p=31$ and controls the rowbuffer reset (line~\ref{line:reset-RB}).  The nanowires are shifted toward AP$_0$ and the value is written back to AP$_1$ (lines~\ref{line:round-robin}--\ref{line:round-robin-end}, see Fig.~\ref{subfig:DBC}) such that after a round of TRD ($=7$ shown in the algorithm) each exponent is written back and stored in its correct original position.  If there are no `1's in this position, the predicate logic is never true because $isAnyOne$ is zero (line~\ref{line:test-ones}), so none of the exponents are eliminated (line~\ref{line:reset-RB}).   

Note, each exponent is logically shifted left by one each time, this is to facilitate reading the predicate from bit $p$ each round.  This ensures the predicate is only selected from position 0, 31, and 47.  The final result is logically shifted right by eight to put the maximum exponent value in the correct alignment (line~\ref{line:shift-right-8}).  The main function uses \textsc{FindMax} in $\log_{\text{TRD}=7}$ fashion to find the overall maximum exponent now stored in E.

\begin{algorithm}[tbp]
\footnotesize
\caption{Floating-point Addition}
\label{alg:fp-add}
\begin{algorithmic}[1]
\State \algorithmicvariables: $M[0..n-1], E[0..n-1], S[0..n-1]$
\State  \Comment {Inputs: $M[0..n-1], E[0..n-1], S[0..n-1]$ Operands}
\State  \Comment {Outputs: $M$ = Mantissa, $E$ = Exponent, $S$ = Sign}
\Statex
\State \Comment{Conduct Floating-Point Addition of $n$ values}
\Function{FPAdd}{M[0..n-1], E[0..n-1], S[0..n-1]}
\State \Comment {Find maximum exponent searching groups of TRD (7)}
\label{line:find-max-exponent}
\State $t \gets \lceil n/7 \rceil; r \gets n\mod7 $
\While {$t>1$}
\For{$i \gets 0:t-1$} 
   \State E[i] $\gets$ \textsc{FindMax}(E[7i..7i+6])
\EndFor
\For{$i \gets 0:r-1$}
    \State E[t+i] $\gets$ E[7*t+i]
\EndFor
\State $t \gets  \lceil (t+r) / 7 \rceil; r \gets (t+r) \mod 7$
\EndWhile
\For{$i \gets r:6$}
    \State E[$i$] $\gets 0$
\EndFor
\State E = \textsc{FindMax}{E[0..6]} 
\label{line:find-max-exponent-end}
\State \Comment {Normalize Mantissas and convert to 2's complement}
\label{line:norm-2s}
\For{$i \gets 0:n-1$}
  \State $m[2*i] = $ \textsc{NormMantissa}$($M[$i$],E,E[$i$]$)$
  \State $m[2*i+1] = $ 0x0
  \label{line:invert-on-sign}
  \State S[$i$] ? $m[2*i] \gets m[2*i]$ \texttt{XOR} 0xFFFFFFFFFFFFFFFF
  \State S[$i$] ? $m[2*i+1] \gets $0x1
\EndFor
\label{line:invert-on-sign-end}
\label{line:norm-2s-end}
\State \Comment {Sum Mantissas}
\label{line:mantissa-reduction}
\State $t \gets \lceil n*2/7 \rceil; r \gets 2*n\mod7 $
\While {$(t > 1) || (r>5)$}
\For{$i \gets 0:t-1$} 
    \State $m[3i...3i+2]\gets$ \textsc{CSA-Reduction}($m[7i..7i+6]$)
    \If {$r>3$} 
      \State $m[3t...3t\!+\!2]\!\!\gets$\textsc{CSA-Reduction}($m[7t..7t\!+\!r]$)
     \Else 
       \State $m[3t...3t+r] \gets $ $m[7t..7t+r]$)
     \EndIf
     \EndFor
\State $t \gets  \lceil (t*3+r) / 7 \rceil; r \gets (t*3+r) \mod 7$
\EndWhile
\label{line:mantissa-reduction-end}
    \State $M \gets $ \textsc{Add}($m[0..r]$,64)
\State $Sum \gets$ \textsc{NormSum}$(M,E)$
\State \Return{E[0]}
\EndFunction
\end{algorithmic}
\end{algorithm}

Next all of the mantissas must be normalized to the maximum exponent and after normalization if their sign is negative inverted in twos complement for summation (Algorithm~\ref{alg:fp-add}, lines~\ref{line:norm-2s}--\ref{line:norm-2s-end}).  The first step uses the \textsc{NormMantissa} helper function (Algorithm~\ref{alg:fp-add-helper}).  First the exponent is subtracted from the maximum exponent (lines~\ref{line:sub-exp}--\ref{line:sub-exp-end}), then each bit is inspected to normalize the mantissa.  Again the predication bit is extracted from the same position $p$ as in the \textsc{FindMax} function.  If the highest two bits are true, the mantissa is shifted by 128 or 64 places which results in setting it to 0x0 (lines~\ref{line:out-of-scope}--\ref{line:out-of-scope-end}).  For the next three bits (5:3) 4, 2, and 1 predicated shifts by 8 are executed (line~\ref{line:pred-shift-8}), followed by 4, 2, and 1 predicted shifts by 1 (line~\ref{line:pred-shift-1}).  Note, these mantissas are still ``normalized'' to bit position 47 not 23 from the multiplication in a prior step, so a shift by 32 is still in scope.  
After normalization, two storage locations for each value are allocated.  If the value is positive the first location gets the mantissa and the second get 0x0.  Using predication from the sign bit (also bit position 31) the mantissa is inverted and the second location is written 0x1 (Algorithm~\ref{alg:fp-add} lines~\ref{line:invert-on-sign}--\ref{line:invert-on-sign-end}).  

\begin{algorithm}[tbp]
\footnotesize
\caption{Addition Helper Functions}
\label{alg:fp-add-helper}
\begin{algorithmic}[1]

\Statex
\State \Comment{Find the Maximum Among 7 values}
\Function{FindMax}{E[0..6],w=8,o=23}
\State $p \gets o + w$
\For{$i \gets o : o + w - 1$}
\State $isAnyOne \gets $ \texttt{OR} E[0..6]  \Comment {TR$\geq1$}
\label{line:test-ones}
     \For{$j \gets 0 : 6$}
        \State $nIsOne \gets $E[j]$_o$ \texttt{XOR} 0xFFFFFFFFFFFFFFFF
        \label{line:test-local-one}
        \State $pred \gets isAnyOne$ \texttt{AND} $nIsOne$
        \label{line:test-local-one-end}
        \State $RB \gets $E[j]$ << 1$

        \State $pred_p $ ? RESET $RB$ \Comment{test at bit $p \implies RB \gets 0$}
        \label{line:reset-RB}
\State SHIFT rows upward \Comment{Shift DWM DBC, reindex E}
\label{line:round-robin}
        \State E[6] $\gets RB$
        \label{line:round-robin-end}
    \EndFor
\EndFor
\State \Return{TR E[0..6] $>> 8$}
\label{line:shift-right-8}
\EndFunction

\Statex
\State \Comment{Normalize a Mantissa based on Exponent Difference}
\Function{NormMantissa}{M, Max, E,w=8,o=23}
\State $p \gets o + w$
\State $t \gets $ E \texttt{XOR} 0xFF800000 \Comment{Invert Exponent}
\label{line:sub-exp}
\State $S \gets $\textsc{Add}$($Max,$t$,0x800000,0,0,w+1,o$)$
\label{line:sub-exp-end}
\For {$i \gets 7:6$}
\label{line:out-of-scope}
  \State $isOne \gets $S_p; $S \gets S << 1$
  \State $isOne$ ? M $\gets$ 0x0
\EndFor
\label{line:out-of-scope-end}
\For {$i \gets 5:0$}
   \State $k \gets i \mod 3$
   \State $isOne \gets $S_p; $S \gets S << 1$
   \For{$j \gets 0:2^k$}
    \If{$i<3$} $isOne$ ? M $\gets$ M $>> 1$
    \label{line:pred-shift-1}
    \Else $\;isOne$ ? M $\gets$ M $>> 8$
    \label{line:pred-shift-8}
    \EndIf
  \EndFor
\EndFor
\State \Return{M}
\EndFunction

\Statex
\State \Comment{Normalize Mantissa based on Summation and prepare Sum}
\Function{NormSum}{M, E}
\State \Comment {Get sign from MSB and if negative, get 2's Complement}\label{line:mantissa-magnitude}
\State $sign \gets $ M
\For{$i \gets 0:3$} \Comment {Move the sign to bit 31}
\State $sign >> 8$
\EndFor
\State $sign_{31}$ ? M $\gets$ M \texttt{XOR} 0xFFFFFFFFFFFFFFFF
\State $sign_{31}$ ? M $\gets$ \textsc{Add}$($M,0x1,0,0,0,64$)$
\label{line:mantissa-magnitude-end}
\State \Comment{Copy mantissa, $<<$ to find first `1', norm exp with}
\State \Comment{predicate, shift mantissa to bit 23}
\State $m \gets $M; $m << 1$; $seenOne \gets $ 0x0
\State \Comment{If Mantissa is higher than bit 47, increase the exp}
\State \Comment{Shift the mantissa right to bit position 47}
\For{$i \gets 15:0$}
  \State $seenThisOne \gets m$
  \texttt{OR} $seenOne$ \label{line:seen-one}
  \State $seenOneFirst \gets seenThisOne$ \texttt{XOR} $seenOne$ \label{line:seen-one-end}
  
  \State $seenOneFirst_{63}$ ? $expAdd \gets i << 23$ 
  \label{line:add-offset}
  \State $seenOne \gets seenThisOne$
  \State $seenOne_{63}$ ? $M \gets M >> 1$ \label{line:before-47}
  \State $m \gets m << 1$
\EndFor
\State \Comment{If Mantissa is lower than bit 47, decrease the exp}
\For{$i \gets 1:47$}
  \State $seenThisOne \gets m$ \texttt{OR} $seenOne$
  \State $seenOneFirst \gets seenThisOne$ \texttt{XOR} $seenOne$
  \State $seenOneFirst_{63}$ ? $expAdd \gets -i << 23$ 
  \State $seenOne \gets seenThisOne$
  \State $nSeenOne\!\gets\!seenOne $ \texttt{XOR} 0xFFFFFFFFFFFFFFFF
    \State \Comment{From 47..24, shift the mantissa right to bit position 23}
  \If{$i < 23$}
  \State $seenOne_{63}$ ? $M \gets M >> 1$ \label{line:before-23}
  \EndIf
    \State \Comment{From 22...0, shift the mantissa left to bit position 23}
  \If{$i > 23$}
  \State $nSeenOne_{63}$ ? $M \gets M << 1$ \label{line:after-23}
  \EndIf
  \State $m \gets m << 1$
\EndFor
\State $exp \gets $ \textsc{ADD}$($E,$expAdd$,0,0,0,8,23$)$ \Comment {Add offset to exp} \label{line:the-end}
\State $sum \gets m$ \texttt{AND} 0x7FFFFF \Comment{Strip leading `1'}
\State $sum \gets sum$ \texttt{OR} $exp$ \texttt{OR} $sign$ \Comment{Recombine}
\State \Return $sum$ \label{line:the-end-end}
\EndFunction
\end{algorithmic}
\end{algorithm}

Once normalized and converted to twos complement form, the mantissas can be summed.  Similar to multiply, \textsc{CSA-Reduction} is used to reduce operands from $7 \rightarrow 3$ until there are $\text{TRD}-2=5$ or fewer which are then summed using \textsc{Add} (lines~\ref{line:mantissa-reduction}--\ref{line:mantissa-reduction-end}).  The last step is to again normalize the resulting mantissa and to reassemble the floating point number.  Normalization is done using \textsc{NormSum} helper function (Algorithm~\ref{alg:fp-add-helper}).  Since this addition used twos complement logic we extract the sign from most significant bit of the full 64-bit cell.  This becomes the predicate to invert and add 0x1 to make the mantissa positive (lines~\ref{line:mantissa-magnitude}--\ref{line:mantissa-magnitude-end}).  We then look for the first `1' in the mantissa in relationship to bit 47.  

The first bit position requires two storage locations, one to store whether a `1' has been seen $seenOne$ and a second $seenThisOne$ which looks for a `1' at this bit position.  $seenThisOne$ is true if we have seen a `1' previously or there is a one in this round (line~\ref{line:seen-one}.  The predicate $seenOneFirst$ comes from \texttt{XOR} which is only true on the first `1' (line~\ref{line:seen-one-end}) and the $expAdd$ value is only set once (line~\ref{line:add-offset}).  Once $seenOne$ is set, we start shifting M to be aligned with bit 23, which requires right shifts if found before bit 23, shown with predicated shifts on $seenOne$ for bits 62:48 (line~\ref{line:before-47}) and 47:24 (line~\ref{line:before-23}).  If $seenOne$ is still not seen by bit 22, we start shifting left governed by the $seenOne$ complement $nSeenOne$ until the `1' is found (line~\ref{line:after-23}) .    The remainder of the function is to combine the normalization exponent offset $expADD$ with E, strip bit 23 and combine the sign, exponent, and mantissa per the floating point standard (lines~\ref{line:the-end}--\ref{line:the-end-end}).  

Note, these algorithms are designed to show the feasibility of the function.  In some cases, optimizations for system performance or code optimizations for expediency may have been excluded to maintain clarity.  For example, while shown here, we can complete the decomposition and recomposition only at the beginning or ending of the full benchmark when communicating with a host processor.  Additionally, while shown for 63-bits, we can use the lower 32 bits from the $sign$ shift first and then work from the $m$ to only pull predicates from position 31 (selecting from three positions, 47, 31, 0).  

The control for these algorithms comes from the hosts/memory controller.  Control of \textbf{for}, \textbf{if}, \textbf{while}, etc. control constructs are governed by the host as they are deterministic and can be entirely unrolled.  Moreover, these can be distributed via single instruction multiple data (SIMD) execution throughout the system (e.g., via different subarrays) for massive parallelism.  Only the predicated instructions use data-based control, and these presume the instructions will be executed or a nop in its place to remain in lock step with the SIMD execution.  Finally, while shown for 32/64 bits, given the row size is 512, 8 items can be packed per row and computed in parallel.

\bibliography{pim}
\bibliographystyle{IEEEtran}